\documentclass[prl,aps,showpacs,twocolumn,longbibliography]{revtex4-1} %

\usepackage{bm}
\usepackage{amsmath}
\usepackage{amssymb}
\usepackage{graphicx}
\usepackage{slashed}
\usepackage{accents}
\usepackage{mathrsfs}
\usepackage{braket}
\usepackage{units}

\usepackage{upgreek}
\usepackage{tensor}
\usepackage{booktabs}
\usepackage[caption=false]{subfig}

\usepackage[colorlinks=true, linkcolor=blue, urlcolor=blue, citecolor=blue, breaklinks=true]{hyperref}
\usepackage{breakurl}

\usepackage[american]{babel}

\DeclareSymbolFont{eulerscript}{U}{eur}{m}{n}
\DeclareSymbolFontAlphabet{\matheuler}{eulerscript}
\DeclareMathAlphabet{\boldgreek}{OML}{zplm}{b}{it}

\newcommand{\dd}{\mathrm{d}}
\newcommand{\ee}{\mathrm{e}}

\newcommand{\Gi}{\mathrm{Gi}}
\newcommand{\Ai}{\mathrm{Ai}}
\DeclareMathOperator{\diag}{diag}
\renewcommand{\Re}{\mathrm{Re}}
\renewcommand{\Im}{\mathrm{Im}}

\newcommand{\nfrac}[2]{{#1}/{#2}}
\newcommand{\abs}[1]{\left|#1\right|}
\renewcommand{\eqref}[1]{Eq.\ (\ref{#1})}
\newcommand{\expectation}[1]{\langle{#1}\rangle}
\newcommand{\mc}[1]{\mathcal{#1}}

\AtBeginDocument{
\heavyrulewidth=.08em
\lightrulewidth=.05em
\cmidrulewidth=.03em
\belowrulesep=.65ex
\belowbottomsep=0pt
\aboverulesep=.4ex
\abovetopsep=0pt
\cmidrulesep=\doublerulesep
\cmidrulekern=.5em
\defaultaddspace=.5em
\tabcolsep=4pt
}

\begin{document}
\title{High-Energy Vacuum Birefringence and Dichroism in an Ultrastrong Laser Field}

\author{Sergey Bragin}

\author{Sebastian \surname{Meuren}}
\email{s.meuren@mpi-hd.mpg.de}
\thanks{now at the Department of Astrophysical Sciences, Princeton University, Princeton, New Jersey 08544, USA}

\author{Christoph H. \surname{Keitel}}

\author{Antonino \surname{Di Piazza}}

\affiliation{Max-Planck-Institut f\"ur Kernphysik, Saupfercheckweg 1, D-69117 Heidelberg, Germany}

\date{\today}
\begin{abstract}
A long-standing prediction of quantum electrodynamics, yet to be experimentally observed, is the interaction between real photons in vacuum. As a consequence of this interaction, the vacuum is expected to become birefringent and dichroic if a strong laser field polarizes its virtual particle--antiparticle dipoles. Here, we derive how a generally polarized probe photon beam is influenced by both vacuum birefringence and dichroism in a strong linearly polarized plane-wave laser field. Furthermore, we consider an experimental scheme to measure these effects in the {nonperturbative} high-energy regime, where the Euler-Heisenberg approximation breaks down. By employing circularly polarized high-energy probe photons, as opposed to the conventionally considered linearly polarized ones, the feasibility of {quantitatively confirming the prediction of nonlinear QED for vacuum birefringence at the $5\sigma$ confidence level} on the time scale of a few days is demonstrated for upcoming $\unit[10]{PW}$ laser systems. Finally, dichroism and anomalous dispersion in vacuum are shown to be accessible at these facilities.
\end{abstract}

\maketitle

In the realm of classical electrodynamics, the electromagnetic field experiences no self-interaction in vacuum \cite{landau_theory_of_fields_1987}. According to quantum electrodynamics (QED), however, a finite photon-photon coupling is induced by the presence of virtual charged particles in the vacuum \cite{landau_quantum_electrodynamics_1982}. For low-frequency electromagnetic fields $F^{\mu\nu}$, such vacuum polarization effects are described by the Euler-Heisenberg Lagrangian density \cite{dunne_heisenbergeuler_2012,dittrich_quantum_vacuum_2000,schwinger_gauge_invariance_1951,heisenberg_folgerungen_1936}. Below the QED critical field $E_{\text{cr}} = \nfrac{m^2}{\abs{e}} \approx \unitfrac[1.3 \times 10^{18}]{V}{m}$, low-frequency vacuum polarization effects are suppressed \cite{king_measuring_2016,battesti_magnetic_2013,di_piazza_review_2012,marklund_nonlinear_effects_2006,ritus_quantum_effects_1985,mitter_quantum_1975} and the density is given by
\begin{gather}
	\label{eq:eulerheisenberg}
	\mc{L}_{\text{EM}} = - \mc{F} + \frac{\alpha}{90\pi E^2_{\text{cr}}} (4\mc{F}^2 + 7\mc{G}^2) + \cdots,
\end{gather}
where $\mc{F} = \nfrac{F_{\mu\nu} F^{\mu\nu}}{4}$ and $\mc{G} = \nfrac{\tilde{F}_{\mu\nu} F^{\mu\nu}}{4}$ are the electromagnetic field invariants %
\footnote{Here, $F^{\mu\nu}$ denotes the electromagnetic field tensor, $\tilde{F}^{\mu\nu} = \nfrac{\epsilon^{\mu\nu\sigma\tau} F_{\sigma\tau}}{2}$ is its dual, and $\epsilon^{\mu\nu\sigma\tau}$ is the totally antisymmetric pseudotensor. Heaviside and natural units are used ($\epsilon_0 = \hbar = c = 1$), i.e., the fine-structure constant is given by $\alpha = \nfrac{e^2}{(4\pi)} \approx \nfrac{1}{137}$; and the metric tensor $g^{\mu\nu} = \diag(1,-1,-1,-1)$ is employed ($m$ and $e<0$ denote the electron mass and charge, respectively).}.

The Euler-Heisenberg Lagrangian predicts that the vacuum resembles a birefringent medium \cite{bialynicka-birula_nonlinear_1970,baier_photon_propagation_1967,klein_birefringence_1964,toll_dispersion_1952}. The smallness of the QED prediction for the light-by-light scattering cross section {in the low-energy regime} opens up the possibility to search for physics beyond the Standard Model, e.g., axionlike or minicharged particles and paraphotons, by measuring {optical vacuum polarization effects \cite{villalba-chavez_minicharged_2016,villalba-chavez_axion-induced_2013,tommasini_precision_2009,abel_illuminating_2008,gies_polarized_2006}, see also \cite{jaeckel_probing_2016,jaeckel_low-energy_2010}}.

Recent astronomical observations hint at the existence of vacuum birefringence \cite{mignani_evidence_for_vacuum_birefingence_2017} (see also the remarks in \cite{capparelli_note_2017,turolla_comment_2017}). However, a direct laboratory-based verification of this fundamental property of the vacuum is still missing. Laboratory experiments like BFRT \cite{cameron_search_1993}, BMV \cite{cadene_bmw_status_2014}, PVLAS \cite{della_valle_pvlas_2016}, and Q\&{}A \cite{chen_qa_experiment_2007} have so far employed magnetic fields to polarize the vacuum and optical photons to probe it, though without reaching the required sensitivity.

The strongest electromagnetic fields of macroscopic extent are nowadays produced by lasers. However, even the intensities $I \sim \unitfrac[10^{23}]{W}{cm^2}$ envisaged for future $\unit[10]{PW}$-class optical lasers \cite{danson_petawatt_lasers_2015,jeong_femtosecond_2014} are still well below the critical intensity $I_{\text{cr}} = E^2_{\text{cr}}  \approx \unitfrac[4.6 \times 10^{29}]{W}{cm^2}$. Therefore, the leading-order correction given in \eqref{eq:eulerheisenberg} is sufficient to describe low-frequency vacuum polarization effects. Recently, various setups have been considered to measure them \cite{shakeri_polarization_2017,schlenvoigt_feasibility_study_2016,karbstein_x-ray_high-intensity_2016,zavattini_polarisation_2016,tennant_four_2016,gies_quantum_2015,fillion-gourdeau_scheme_2015,karbstein_photon_2015,hu_modified_2014,mohammadi_circular_2014,monden_interaction_2012,king_photonphoton_2012,kryuchkyan_bragg_2011,homma_probing_2011,king_matterless_double_slit_2010,tommasini_detecting_2008,lundstrom_photon-photon_2006,di_piazza_light_diffraction_2006,heinzl_vacuum_birefringence_2006}, but all suggested experiments will remain challenging in the foreseeable future. 

As the light-by-light scattering cross section attains its maximum at the pair-production threshold \cite{landau_quantum_electrodynamics_1982}, it is natural to consider high-energy photons to probe vacuum birefringence \cite{king_vacuum_birefringence_2016,ilderton_prospects_2016,nakamiya_probing_2017,dinu_helicity_flip_2014,wistisen_vacuum_birefringence_2013,cantatore_vacuum_birefringence_1991}. A photon four-momentum $q^{\mu}$ ($q^0 = \omega$, $q^2=0$) allows us to construct a third invariant, the quantum nonlinearity parameter (see \cite{landau_quantum_electrodynamics_1982}, \S~101)
\begin{gather}
	\label{eq:chi}
	\chi = \frac{\sqrt{-(f^{\mu\nu}q_{\nu})^2}}{E_{\text{cr}} m} \approx 0.5741 \, \frac{\omega}{\unit{GeV}}\, \sqrt{ \frac{I}{\unitfrac[10^{22}]{W}{cm^2}}}
\end{gather}
[for a plane-wave background field with amplitude $f^{\mu\nu}$ {and phase-dependent pulse shape $\psi'(\phi)$, i.e., $F^{\mu\nu} = f^{\mu\nu} \psi'(\phi)$, details are given below}; the last relation in \eqref{eq:chi} assumes a head-on collision]. As gamma photons with energies $\omega \gtrsim \unit[1]{GeV}$ are obtainable from Compton backscattering \cite{muramatsu_spring-8_2014,weller_higammas_2009,fukuda_circularly_polarized_gammas_2003,ginzburg_colliding_1984,landau_quantum_electrodynamics_1982}, the regime $\chi \sim 1$ is attainable in future laser-based vacuum birefringence experiments.

\begin{figure}
	\centering
	\begin{minipage}{\columnwidth}
	\includegraphics[scale=0.67]{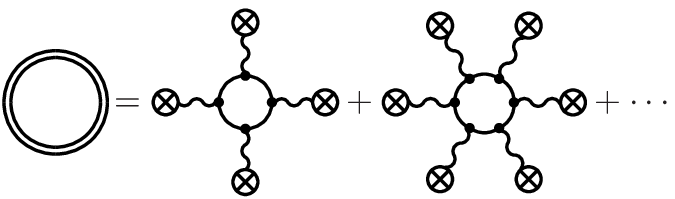} \hspace*{2pt} \raisebox{-1pt}{%
	\includegraphics[scale=0.67]{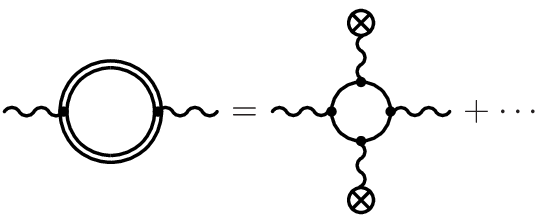}}
	\end{minipage}
	\begin{minipage}{\columnwidth}
	\raisebox{-12pt}{\hspace*{10pt}
	\textbf{(a)} \small Euler-Heisenberg
	\hspace*{30pt}
	\textbf{(b)} \small Polarization operator}
	\end{minipage}
	\caption{\label{fig:loops}The Euler-Heisenberg effective action is only valid for approximately constant fields (denoted by a wiggly line with a cross). The polarization operator must be considered if the momentum of the probe photon (wiggly line) becomes influential ($\chi \gtrsim 1$). Here, a solid double line denotes the exact electron propagator inside a classical background field.}
\end{figure}

In the {nonperturbative} regime $\chi \gtrsim 1$ the Euler-Heisenberg approximation is no longer applicable, as it neglects the contribution of the probe photon momentum which flows in the electron-positron loop (see Fig.~\ref{fig:loops}a). Instead, the polarization operator in the background field must be employed (see Fig.~\ref{fig:loops}b). For low-energy photons, both objects in Fig.~\ref{fig:loops} are related by functional derivatives \cite{bialynicka-birula_nonlinear_1970}. The regime $\chi \gtrsim 1$ is qualitatively different from the one where the Euler-Heisenberg approximation is valid, in particular, due to the following two reasons: 1) electron-positron photoproduction becomes sizable, and thus, the vacuum acquires dichroic properties; 2) the vacuum exhibits anomalous dispersion \cite{heinzl_exploring_2009,dinu_helicity_flip_2014,ritus_quantum_effects_1985,baier_polarization_operator_1976,becker_vacuum_1975}.

In this Letter, we put forward an experimental scheme to measure high-energy vacuum birefringence and dichroism in the nontrivial regime $\chi \gtrsim 1$. It is based on Compton backscattering to produce polarized gamma photons and exploits pair production in matter to determine the polarization state of the probe photon after it has interacted with a {linearly polarized} strong laser pulse. By analyzing the consecutive stages of this type of experiment, we show that for vacuum birefringence, the required measurement time is reduced by two orders of magnitude if a circularly polarized probe photon beam is employed (hitherto, only linearly polarized probe gamma photons have been considered for setups similar to ours \cite{king_vacuum_birefringence_2016,ilderton_prospects_2016,nakamiya_probing_2017,dinu_helicity_flip_2014},%
\footnote{So far, the case of circularly polarized photons has only been discussed for proposals to measure vacuum birefringence in magnetic fields at $\chi\ll 1$ \cite{wistisen_vacuum_birefringence_2013,cantatore_vacuum_birefringence_1991}.}).

Assuming conservative experimental parameters, we demonstrate that with this type of setup and the observables we introduce [see \eqref{eq:rb_experimental}], the quantitative verification of the strong-field QED prediction for vacuum birefringence and dichroism is feasible with an average statistical significance of $5\sigma$ on the time scale of a few days at upcoming $\unit[10]{PW}$ laser facilities. 

In the following, we consider a linearly polarized plane-wave laser pulse, described by the four-potential $A^{\mu}(kx) = a^{\mu} \psi(kx)$. Here, $x^\mu$ is the position four-vector, $k^{\mu}$ is a characteristic laser photon four-momentum ($k^0 = \omega_\textrm{L}$, $k^2 = 0$), $a^{\mu}$ characterizes the amplitude of the field ($a^2 < 0$, $ka = 0$, $f^{\mu\nu} = k^{\mu}a^{\nu} - k^{\nu}a^{\mu}$), and $\psi(kx)$ defines its pulse shape ($|\psi(kx)|,\;|\psi'(kx)| \lesssim 1$; a prime denotes the derivative of a function with respect to its argument).

A gauge- and Lorentz-invariant measure of the laser field strength is the classical intensity parameter \cite{ritus_quantum_effects_1985}
\begin{equation}
	\label{eq:xi}
	\xi = \frac{|e| \sqrt{-a^2}}{m} \approx 0.7495 \, \frac{\unit{eV}}{\omega_\textrm{L}} \sqrt{ \frac{I}{\unitfrac[10^{18}]{W}{cm^2}}}.
\end{equation}
Here, we focus on high-intensity optical lasers ($I \gtrsim \unitfrac[10^{20}]{W}{cm^2}$, $\omega_\textrm{L} \sim \unit[1]{eV}$), i.e., the regime $\xi \gg 1$.

\begin{figure}
	\centering
	\begin{minipage}{\columnwidth}
		\includegraphics[scale=0.7]{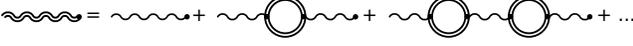}
	\end{minipage}%
	\caption{A background field changes the photon dispersion relation via radiative corrections induced by virtual particles \cite{landau_quantum_electrodynamics_1982}. Here, we neglect higher-order radiative corrections to the electron-positron loop as $\alpha\chi^{\nfrac{2}{3}} \ll 1$  \cite{ritus_quantum_effects_1985}.}
	\label{fig:exact_photon}
\end{figure}

Inside a plane-wave background field an incoming external photon line (see Fig.~\ref{fig:exact_photon}) in a Feynman diagram corresponds (up to normalization) to the function $\Phi_q^{\mu}(x)$, which is a solution of the Dyson equation~\cite{meuren_pair_creation_2015,landau_quantum_electrodynamics_1982} with initial condition $\Phi_q^{\mu}(x) \to \Phi_q^{(0)\mu}(x) = \epsilon^{(0)\mu} \ee^{-iqx}$ as $kx \to -\infty$ ($\epsilon^{(0)}\epsilon^{(0)*} = -1$, $q\epsilon^{(0)} = 0$). After applying the local constant field approximation (valid if $\xi\gg 1$) and following \cite{meuren_pair_creation_2015}, we find that to leading order, $\Phi_q^{\mu}(x)$ is given by (see also \cite{villalba-chavez_minicharged_2016,dinu_helicity_flip_2014,baier_polarization_operator_1976,becker_vacuum_1975})

\begin{equation}
	\label{eq:photon_inside_the_pulse}
	\Phi_q^{\mu}(x) = \epsilon^{\mu}(kx) \ee^{-iqx},
	\quad
	\epsilon^{\mu}(kx) = \sum_{i=1,2} c\indices*{_{i}}(kx) \Lambda_i^\mu,
\end{equation}
where
\begin{equation}
	\label{eq:freepolvecexpansion}
	\epsilon^{\mu}(kx\to-\infty) = \epsilon^{(0)\mu} = \sum_{i=1,2} c\indices*{_{i}^{(0)}} \Lambda_i^\mu,
\end{equation}
and $\Lambda_1^\mu = {f^{\mu\nu} q_\nu}/{\sqrt{qf^2q}}$, $\Lambda_2^\mu = - {\tilde{f}^{\mu\nu} q_\nu}/{\sqrt{qf^2q}}$ ($q\Lambda_i = k\Lambda_i = 0$, $\Lambda_i \Lambda_{j} = -\delta_{ij}$; note that $\Lambda_2^\mu$ is actually a pseudo four-vector)~\cite{meuren_high-energy_2015,meuren_pair_creation_2015, meuren_polarization_operator_2013}.
The coefficients $c\indices*{_{i}}(kx)$ and $c\indices*{_{i}^{(0)}}$ are connected via
\begin{equation}
	\label{eq:dyson_equation_solution}
    c\indices*{_{i}}(kx) = c\indices*{_{i}^{(0)}} \exp{[i\phi_i(kx) -\lambda_i(kx)]},
\end{equation}
where
\begin{equation}
	\label{eq:shift_and_decay_inside_the_pulse}
	\begin{bmatrix} \phi_i(kx) \\ \lambda_i(kx) \end{bmatrix} =  - \frac{1}{2kq} \int\displaylimits_{-\infty}^{kx} \dd \phi \, \begin{bmatrix} \Re \left[ p_i(\phi,\chi) \right] \\ \Im \left[ p_i(\phi,\chi) \right] \end{bmatrix},
\end{equation}
[we refer to $\phi_i = \phi_i(kx\to \infty)$ as phase shifts and to $\lambda_i = \lambda_i(kx\to \infty)$ as decay parameters] %
with
\begin{equation}
	\label{eq:polarization_operator_coefficients}
	\begin{bmatrix} p_1(kx,\chi) \\ p_2(kx,\chi) \end{bmatrix} 
		= 
	\frac{\alpha m^2}{3 \pi} \int\displaylimits_{-1}^{+1} \dd v \begin{bmatrix} (w - 1) \\ (w + 2) \end{bmatrix}   \frac{f'(u)}{u},
\end{equation}
$w = 4/(1-v^2)$, $u = [w/\chi(kx)]^{2/3}$, $\chi(kx) = \chi |\psi'(kx)|$, and $f(u) = \pi \left[ \Gi(u) + i \Ai(u) \right]$ \cite{olver_nist_handbook_2010,ritus_quantum_effects_1985}.

\begin{figure*}
	\begin{minipage}[T]{0.62\textwidth}
		\raisebox{70pt}{\small\textbf{(a)}}\hspace*{-2pt}
		\includegraphics[scale=1]{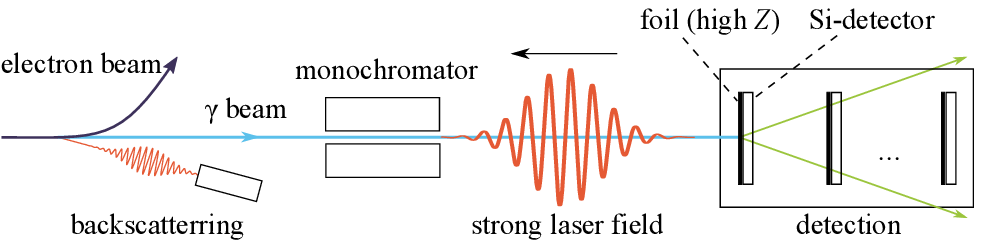}
	\end{minipage}
	\begin{minipage}[T]{0.36\textwidth}
		\raisebox{70pt}{\small\textbf{(b)}}\hspace*{-2pt}
		\includegraphics[scale=1]{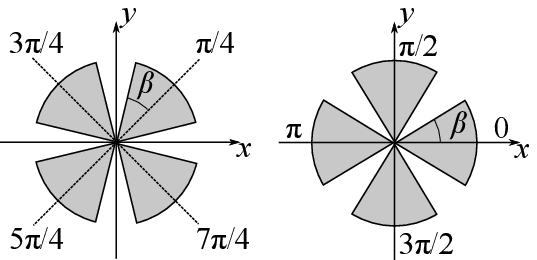}
	\end{minipage}
	\caption{\textbf{(a)} Experimental setup. Polarized highly energetic gamma photons (produced via Compton backscattering) propagate through a strong laser field, which induces vacuum birefringence and dichroism. Afterward, the gamma photons are converted into electron-positron pairs. From their azimuthal distribution, the polarization state is deduced.  \textbf{(b)} Regions of the transverse plane (gray), which are used to define the observables $R_\textrm{B}$ (left) and $R_\textrm{D}$ (right) [see \eqref{eq:rb_experimental}].}
	\label{fig:experiment}
\end{figure*}

In order to extend the above result from a single photon to a photon beam (which is, in general, not in a pure polarization state), we introduce the following density tensors, which describe the initial ($\varrho^{(0)\mu \nu}$) and the final ($\varrho^{\mu \nu}$) polarization state of the beam \cite{landau_quantum_electrodynamics_1982,meuren_semiclassical_2016,blum_density_matrix_2012}
\begin{equation}
	\begin{aligned}
	\label{eq:density_tensors_definition}
		 {\varrho}^{(0)\mu \nu} &=  \sum_{a} w\indices{_a} \epsilon\indices*{_a^{(0)\mu}} \epsilon\indices*{_a^{(0)*\nu}} &=& \sum_{i,j=1,2} \rho\indices*{_{ij}^{(0)}} \Lambda_i^{\mu} \Lambda_j^{\nu},\\
		 {\varrho}^{\mu \nu} &=  \sum_{a} w\indices{_a} \epsilon\indices*{_a^{\mu}} \epsilon\indices*{_a^{*\nu}} &=& \sum_{i,j=1,2} \rho\indices*{_{ij}} \Lambda_i^{\mu} \Lambda_j^{\nu}.
	\end{aligned}
\end{equation}
Here, $w\indices{_a}$ represents the probability to find a photon with polarization four-vector $\epsilon\indices*{_a^{(0)\mu}}$ ($\epsilon\indices*{_a^{\mu}}$) in the initial (final) beam.

Using the identity matrix $\mathrm{I}$ and the Pauli matrices $\boldsymbol{\sigma} = \left( \sigma_1, \sigma_2, \sigma_3 \right)$ \cite{landau_quantum_electrodynamics_1982}, we expand the initial ($\rho\indices*{_{ij}^{(0)}}$) and the final ($\rho\indices*{_{ij}}$) polarization density matrices as \cite{landau_quantum_electrodynamics_1982,meuren_semiclassical_2016,blum_density_matrix_2012}
\begin{equation}
	\label{eq:density_matrix_expansion}
	\rho^{(0)} = \frac{1}{2} \left(S_0^{(0)}\mathrm{I} + \boldsymbol{S}^{(0)}\boldsymbol{\sigma} \right), 
	\quad
   \rho = \frac{1}{2} \left(S_0\mathrm{I} + \boldsymbol{S}\boldsymbol{\sigma} \right)
\end{equation}
[$\mathrm{Tr}(\rho^{(0)}) = S_0^{(0)}$, $\mathrm{Tr}(\rho) = S_0$; $S_0 \leq S_0^{(0)}$, in general, as the photons can decay in the strong background field]. 

The real Stokes parameters $S^{(0)} = \{S_0^{(0)}, \boldsymbol{S}^{(0)}\}$ [$\boldsymbol{S}^{(0)}=(S_1^{(0)}, S_2^{(0)}, S_3^{(0)})$] and $S = \{S_0, \boldsymbol{S}\}$ [$\boldsymbol{S}=\left( S_1, S_2, S_3 \right)$] completely characterize the initial (final) polarization state of the beam \cite{blum_density_matrix_2012,born_wolf_optics_1999}. Therefore, the following relations describe any possible vacuum birefringence and/or dichroism experiment [see Eqs.~(\ref{eq:photon_inside_the_pulse}), (\ref{eq:dyson_equation_solution}), (\ref{eq:density_tensors_definition}), and (\ref{eq:density_matrix_expansion})]
\begin{gather}
\label{eq:stokes_parameters_connection}
	\begin{aligned}
	\begin{pmatrix}
	S_0 \\ S_3 
	\end{pmatrix}
	&=
	 \ee^{-(\lambda_1 + \lambda_2)}
	\begin{pmatrix}
	\cosh\delta\lambda &  \sinh\delta\lambda
	\\
	\sinh\delta\lambda &  \cosh\delta\lambda
	\end{pmatrix}
	\begin{pmatrix}
	S_0^{(0)} \\ S_3^{(0)} 
	\end{pmatrix},
	\\
	\begin{pmatrix}
	S_1 \\ S_2  
	\end{pmatrix}
	&=
	 \ee^{-(\lambda_1 + \lambda_2)}
	\begin{pmatrix}
	\cos\delta\phi & -\sin\delta\phi
	\\
	\sin\delta\phi & \phantom{+}\cos\delta\phi
	\\
	\end{pmatrix}
	\begin{pmatrix}
	S_1^{(0)} \\ S_2^{(0)}
	\end{pmatrix}.
	\end{aligned}
\end{gather}
Here, $\delta\phi = \phi_2 - \phi_1$ is related to vacuum birefringence and $\delta\lambda = \lambda_2 - \lambda_1$ to vacuum dichroism.

In the following, we discuss possible high-energy vacuum birefringence and/or dichroism experiments (see Fig.~\ref{fig:experiment}a) at the Apollon facility (F1/F2 laser)~\cite{papadopoulos_apollon_2016}, ELI-NP (two $\unit[10]{PW}$ lasers) \cite{negoita_eli-np_2016,turcu_high_2016}, and  ELI-Beamlines (ELI-BL; L3/L4 laser) \cite{rus_eli-beamlines_laser_systems_2013}. At each facility, a 10 PW laser is employed to polarize the vacuum and the second laser is utilized to produce electron bunches via laser wakefield acceleration~\cite{leemans_multi-gev_electron_beams_2014,wang_laser-plasma_acceleration_2013}. We also consider a possible experiment (denoted as LINAC-L) at a conventional electron accelerator, e.g., the European XFEL~\cite{euroxfel}, FACET-II~\cite{facet-ii_website}, or SACLA~\cite{yabashi_overview_2015}, combined with a high-repetition (10 Hz) 1 PW laser. The parameters of the considered facilities are summarized in the Supplemental Material~%
\footnote{\label{supplement}See Supplemental Material for technical details, which includes Refs.~\cite{akhiezer_quantum_electrodynamics_1969,le_garrec_eli-beamlines_2014,eli-beamlines_lasers_website,tsai_pair_production_1974,riley_mathematical_2006,james_statistical_2006,ku_notes_1966}}.

{We assume that $N_e = 10^8$ monoenergetic few-$\unit{GeV}$ electrons are used in one experimental cycle for the generation of probe gamma photons via Compton backscattering.}

\begin{figure}[b!]
	\centering
	\includegraphics[scale=1]{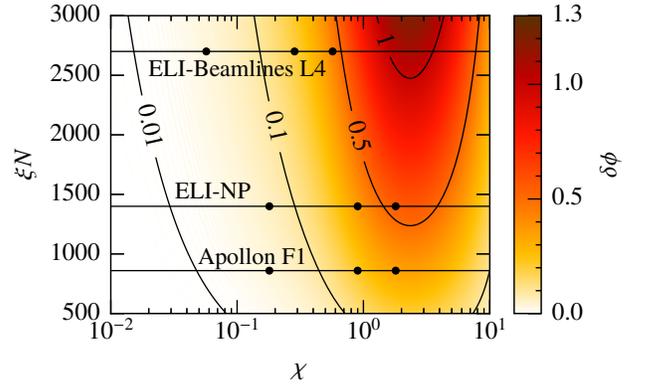}
	\caption{Plot of $\delta\phi$ as a function of $\chi$ and $\xi N$ for a rectangular pulse profile. For each of the three laser facilities, gamma photons with energy $\omega = \unit[0.1]{GeV}$ (left point), $\omega = \unit[0.5]{GeV}$ (central point), and $\omega = \unit[1]{GeV}$ (right point) are indicated. {Note that $|\delta\phi| \gtrsim 0.1$ is also achievable by employing a longer PW laser pulse (e.g., National Ignition Facility with $\Delta t = \unit[3]{ns}$~\cite{danson_petawatt_lasers_2015}) and probe photons with $\omega \gtrsim \unit[0.1]{GeV}$.}}
	\label{fig:rectangular_chi_xi_n}
\end{figure}

For a rectangular pulse with $N$ cycles \{$\psi'(kx) = \sin(kx)$ if $kx \in [-N\pi, N\pi]$ and $\psi'(kx) = 0$ otherwise\}, the relative phase shift $\delta\phi$ depends only on $\chi$ and $\xi N$; it is plotted in Fig.~\ref{fig:rectangular_chi_xi_n}. We conclude that $\abs{\delta\phi} \lesssim 0.1$ for upcoming laser systems in the regime $0.1 \lesssim \chi < 1$, where a clean vacuum birefringence measurement is feasible as pair production is exponentially suppressed. Notably, the quantity $\delta\phi$ decreases with the increase of the probe photon energy for $\chi \gtrsim 2.5$, which characterizes the anomalous dispersion of the vacuum in this regime~\cite{heinzl_exploring_2009,dinu_helicity_flip_2014,ritus_quantum_effects_1985,baier_polarization_operator_1976,becker_vacuum_1975}.

For obtaining better estimates as those given in Fig.~\ref{fig:rectangular_chi_xi_n}, in the following, we employ a Gaussian pulse envelope $\psi'(kx) = \exp[-(\nfrac{kx}{\Delta\phi})^2] \sin (kx)$, where $\Delta\phi$ is related to the duration of the pulse $\Delta t$ (FWHM of the intensity) via $\Delta\phi = \omega_{\textrm{L}}\Delta t / \sqrt{2\ln2}$. This  pulse collides with $N_\gamma = N_e \sigma_\textrm{bs} (\nfrac{I_\textrm{bs}}{\omega_\textrm{bs}}) \Delta t_\textrm{bs}$ gamma photons, where $\sigma_\textrm{bs}$ is the cross section of Compton scattering \cite{landau_quantum_electrodynamics_1982}, and the index ``bs'' indicates the parameters characterizing the backscattering process. To obtain a high degree of polarization, we consider only photons which are scattered in the region $\theta \in (0, \theta_\textrm{max}\ll 1)$, where $\theta$ denotes the polar angle ($\theta = 0$ corresponds to perfect backscattering) \cite{muramatsu_spring-8_2014,weller_higammas_2009,fukuda_circularly_polarized_gammas_2003,ginzburg_colliding_1984,landau_quantum_electrodynamics_1982},\cite{Note3}.  

Below, we employ $\Delta t_\textrm{bs} = \Delta t$, $\omega_\textrm{bs} = \unit[1.55]{eV}$, and $I_\textrm{bs} = \unitfrac[4.3 \times 10^{16}]{W}{cm^2}$ [considering linear Compton scattering is sufficient as $\xi_\textrm{bs} = 0.1$ for this laser; see \eqref{eq:xi}]. 

One of the main experimental challenges is to analyze the final polarization state of the gamma photons. Here, we consider pair production in a screened Coulomb field of charge $Z|e|$ \cite{hunter_pair_production_telescope_2014,bernard_polarimetry_2013,kelner_methods_1975,olsen_pair_production_1959}. The spin-summed pair production cross section is given by
\begin{equation}
	\label{eq:pair_production_coulomb_stokes}
	\dd \sigma_\textrm{pp} = \frac{\dd \varphi}{2 \pi} \big\{ S_0 \sigma_0 + [S_1 \sin(2\varphi) + S_3 \cos(2\varphi)] \sigma_1 \big\},
\end{equation}
where $\varphi$ denotes the azimuth angle of the electron momentum in the transverse plane. For $\sigma_0$, $\sigma_1$, we use expressions exact in $Z\alpha$ and valid for ultrarelativistic particles~\cite{kelner_methods_1975,olsen_pair_production_1959},\cite{Note3}. In the following, we assume a head-on collision [$q^\mu = \omega(1, 0, 0, 1)$, $k^\mu = \omega_{\textrm{L}} (1,0,0,-1)$, $\Lambda_1^\mu = (0,1,0,0)$, $\Lambda_2^\mu = (0,0,1,0)$], and tungsten ($Z=74$) as conversion material. %

As the pair-production cross section is only sensitive to linear polarization [$S_1$ and $S_3$, see \eqref{eq:pair_production_coulomb_stokes}], we conclude from \eqref{eq:stokes_parameters_connection} that we need to utilize circularly polarized probe photons (e.g., $S^{(0)} = \{1, 0, -1, 0\}$) in order to obtain probabilities which depend on $\delta\phi$ [rather than $(\delta\phi)^2$] if $\abs{\delta\phi} \ll 1$ (see also \cite{wistisen_vacuum_birefringence_2013,cantatore_vacuum_birefringence_1991}). Therefore, inverting the standard scheme by using circularly instead of linearly polarized probe photons is highly beneficial in the regime $\abs{\delta\phi} \lesssim 0.1$. 

\begin{table}
\begin{center}
	\begin{tabular}{cccccccccc}
	\toprule
	& $1-S_0$ & $S_1$ & $\expectation{R_\textrm{B}}$ & $N\indices*{_\gamma^{\textrm{B}}}$ &  $\tau$\\
	\midrule
	Apollon     & $1.9 \times 10^{-5}$ & 0.06 & $3.4 \times 10^{-3}$ & $3.0 \times 10^8$    & {45 d}\\
	ELI-NP  & $3.1 \times 10^{-5}$ & 0.09 & $5.6 \times 10^{-3}$ & $1.1 \times 10^8$    & {10 d}\\
	ELI-BL  & $6.3 \times 10^{-5}$ & 0.18 & $1.1 \times 10^{-2}$ & $2.6 \times 10^7$    & {11 h}\\
	{LINAC-L} & $3.8 \times 10^{-6}$ & 0.01 & $6.8 \times 10^{-4}$ & $7.4 \times 10^{9}$ & {2 d}\\
	\bottomrule
	\end{tabular}
	\caption{\label{tab:examples} Duration of the experiment $\tau$ at different facilities ($\chi = 0.25$). $S_0$ and $S_1$, $\expectation{R_\textrm{B}}$, and $N\indices*{_\gamma^{\textrm{B}}}$ follow from \eqref{eq:stokes_parameters_connection}, \eqref{eq:rb_rd_calculated} and \eqref{eq:nb}, respectively ($S^{(0)} = \{1, 0, -1, 0\}$; $5\sigma$ confidence level, i.e., $n=5$). Note that the pair production probability in the strong laser field is much smaller than the conversion efficiency in the detector [$(1 - S_0) \ll \eta = 10^{-2}$].}
\end{center}
\end{table}

From \eqref{eq:stokes_parameters_connection}, we conclude that $S_1$ is sensitive to vacuum birefringence ($\delta\phi$), whereas $S_3$ depends on vacuum dichroism ($\delta\lambda$). To disentangle both effects, we introduce the following asymmetries:
\begin{gather}
	\label{eq:rb_experimental}
	\begin{aligned}
	R_\textrm{B} &= \frac{(N_{\nfrac{\pi}{4}} + N_{\nfrac{5\pi}{4}}) - (N_{\nfrac{3\pi}{4}} + N_{\nfrac{7\pi}{4}})}{(N_{\nfrac{\pi}{4}} + N_{\nfrac{5\pi}{4}}) + (N_{\nfrac{3\pi}{4}} + N_{\nfrac{7\pi}{4}})},
\\
	R_\textrm{D} &= \frac{(N_{0} + N_{\pi}) - (N_{\nfrac{\pi}{2}} + N_{\nfrac{3\pi}{2}})}{(N_{0} + N_{\pi}) + (N_{\nfrac{\pi}{2}} + N_{\nfrac{3\pi}{2}})},
	\end{aligned}
\end{gather}
where $N_{\beta_0}$ denotes the number of pairs detected in the azimuth angle range $\varphi \in (\beta_0 - \beta, \beta_0 + \beta)$ of the transverse plane, with $\beta$ being specified below (see Fig.~\ref{fig:experiment}b). The  expectation values of $R_\textrm{B}$ and $R_\textrm{D}$ are given by [see \eqref{eq:pair_production_coulomb_stokes}] 
\begin{equation}
	\label{eq:rb_rd_calculated}
	\expectation{R_\textrm{B}} = \frac{\sin(2\beta)}{2\beta} \, \frac{\sigma_1}{\sigma_0} \, \frac{S_1}{S_0},
	\quad
	\expectation{R_\textrm{D}} = \frac{\sin(2\beta)}{2\beta} \, \frac{\sigma_1}{\sigma_0} \, \frac{S_3}{S_0}.
\end{equation}

In order to detect vacuum birefringence (dichroism) at the $n\sigma$ confidence level on average, we require that the expectation value $\expectation{R_\textrm{B}}$ ($\expectation{R_\textrm{D}}$) differs from zero by $n$ standard deviations. Therefore, we obtain the following expressions for the number of required incoming gamma photons (see Supplemental Material~\cite{Note3}):
\begin{equation}
	\label{eq:nb}
	N\indices*{_\gamma^{\textrm{B}}} = \frac{\pi n^2}{4\eta \beta S_0 \expectation{R_\textrm{B}}^2},
	\quad
	N\indices*{_\gamma^{\textrm{D}}} = \frac{\pi n^2}{4\eta \beta S_0 \expectation{R_\textrm{D}}^2}
\end{equation}
[by minimizing $N\indices*{_\gamma^{\textrm{B}}}$ ($N\indices*{_\gamma^{\textrm{D}}}$), we find the optimal angle $\beta = \beta_{\textrm{opt}} \approx 0.58 \approx 33^{\circ}$ for both observables]. Here, $\eta = n_z l \sigma_0$ denotes the photon to pair conversion efficiency ($n_z$ and $l$ are the number density and the thickness of the conversion material, respectively). The thickness of a conversion foil should be $\lesssim 1$ milliradiation length (mRL), otherwise multiple Coulomb scattering affects the measured angle \cite{hunter_pair_production_telescope_2014,kelner_methods_1975}. Supposing that several conversion foils alternating with silicon detectors are cascaded \cite{peitzmann_prototype_2013,fermi_lat_2009,agile_2003}, we assume here $\eta = 10^{-2}$ (i.e., an effective thickness of $\sim\unit[10]{mRL}$).

To obtain a clean vacuum birefringence experiment without real electron-positron pair production, we consider the case $\chi = 0.25$. The results for the {four} facilities under consideration are summarized in Table~\ref{tab:examples}. As expected from Fig.~\ref{fig:rectangular_chi_xi_n}, ELI-Beamlines is the most suitable facility for carrying out the measurement in this regime (the expected measurement time is less than one day).

As the number of required gamma photons $N\indices*{_\gamma^{\textrm{B}}}$ scales as $\expectation{R_\textrm{B}}^{-2}$ [see \eqref{eq:nb}], the use of circularly polarized probe photons instead of linearly polarized ones reduces the measurement time by a factor $\approx 100$ ($\delta\phi \approx 0.1$, see Fig.~\ref{fig:rectangular_chi_xi_n}). 

\begin{figure}
	\centering
	\includegraphics[scale=1.0]{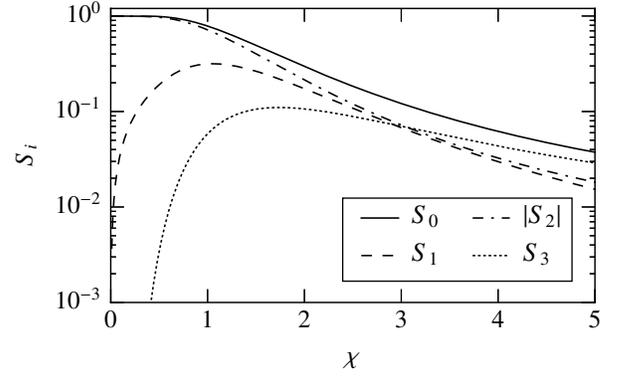}
	\caption{Final Stokes parameters [see \eqref{eq:stokes_parameters_connection}] for gamma photons propagating through an ELI-NP $\unit[10]{PW}$ laser pulse ($S^{(0)} = \{1, 0, -1, 0\}$). The strongest effect is obtained around $\chi = 1$ {(note that pair production becomes sizable for $\chi \gtrsim 1$). As we consider the tunneling regime $\nfrac{1}{\xi} \ll 1$, cusplike structures -- characteristic for multiphoton pair production \cite{villalba-chavez_minicharged_2016,becker_vacuum_1975} -- are absent.}}
	\label{fig:stokes}
\end{figure}

Finally, we consider the case $\chi = 2.5$ (attainable, e.g., at ELI-NP by utilizing $\unit[8.4]{GeV}$ electrons for backscattering; $\theta_{\textrm{max}} = 7.6 \times 10^{-6}$, $\sigma_{\textrm{bs}} = 0.135r_e^2$, $\omega = 1.4$~GeV, $\sigma_1/\sigma_0 = 0.077$; $r_e = \alpha/m = 2.818 \times 10^{-13}$~cm is the classical electron radius). In this regime, vacuum dichroism and anomalous dispersion come into play and the Euler-Heisenberg approximation breaks down completely (see Fig.~\ref{fig:rectangular_chi_xi_n}), {whereas the production of particles, heavier than electrons and positrons, and QCD corrections are still suppressed~\cite{bern_qcd_2001}}. As the produced pairs radiate gamma photons, a discrimination of primary from secondary photons is necessary, e.g., via determination of the photon energy. For $S^{(0)} = \{1, 0, -1, 0\}$, we obtain that $S = \{0.18, 0.11, -0.12, 0.09\}$ at ELI-NP (see Fig.~\ref{fig:stokes}). Correspondingly, $\expectation{R_\textrm{B}} = 3.6 \times 10^{-2}$ and $\expectation{R_\textrm{D}} = 3.0 \times 10^{-2}$, implying a measurement time of {3-4 days [$5\sigma$ confidence level, see \eqref{eq:nb}]}.

\begin{acknowledgments}
	We would like to thank Oleg Skoromnik for useful discussions and Silvia Masciocchi for useful comments on the detection of gamma photons. S. M. was partially supported by the German Research Foundation (Deutsche Forschungsgemeinschaft, DFG) -- ME 4944/1-1.
\end{acknowledgments}

\end{document}

\def\theequation{S\arabic{equation}}

\title{Supplementary material}

\title{High-Energy Vacuum Birefringence and Dichroism in an Ultrastrong Laser Field: Supplemental Material}

\author{Sergey Bragin}

\author{Sebastian \surname{Meuren}}

\author{Christoph H. \surname{Keitel}}

\author{Antonino \surname{Di Piazza}}

\affiliation{Max-Planck-Institut f\"ur Kernphysik, Saupfercheckweg 1, D-69117 Heidelberg, Germany}

\maketitle

The supplemental material is organized as follows: in \secref{sec:lasers} the laser parameters are summarized, in \secref{sec:compton} Compton backscattering is discussed, in \secref{sec:pair} pair production in a Coulomb field is reviewed, the statistical analysis is presented in \secref{sec:statistics}, and, finally, the technical parameters of a potential linear accelerator based experiment are considered in \secref{sec:accelerator} [note that in \secref{sec:compton}, including Table~\ref{tab:compton}, the notation differs partly from the one in the rest of the supplemental material and the main text of the paper]. Even though all given expressions are easily obtainable from those published in the cited literature, we provide them here for the convenience of the reader.

\section{Laser parameters}
\label{sec:lasers}

In Table~\ref{tab:facilities} the parameters of the ultrahigh-intensity lasers, which are considered in the numerical calculations, are shown [photon energy $\omega_\textrm{L}$, pulse energy $\mathcal{E}$, pulse duration $\Delta t$, peak focused intensity $I$, and pulse repetition rate (PRR)]. From them we deduce $\xi$, $\chi$, the number of cycles $N$ and the pulse width $\Delta \phi$ used for the Gaussian envelope.

\section{Compton backscattering}
\label{sec:compton}

Our discussion of linear Compton scattering closely follows \cite{landau_quantum_electrodynamics_1982} \S~86/87 (see also \cite{akhiezer_quantum_electrodynamics_1969} and \cite{ginzburg_colliding_1984}).

The four vectors $p^\mu = (\epsilon, \boldsymbol{p})$ and $k^\mu = (\omega_\text{bs}, \boldsymbol{k})$ [$p'^\mu = (\epsilon', \boldsymbol{p}')$ and $q^\mu = (\omega, \boldsymbol{q})$] denote the four-momenta of the initial [final] electron and photon, respectively. We assume a head-on collision and direct the $z$-axis along the initial electron momentum $\boldsymbol{p}$ [$p^\mu = (\epsilon, 0, 0, p_{z}), k^\mu = \omega_\text{bs}(1, 0, 0, -1)$].

We consider an unpolarized incoming electron beam and sum over the polarization of the outgoing electrons. The polarization state of the initial photon beam and the state selected by the detector, which measures the final photon polarization, are described by the density tensors $\varrho^{\mu\nu}$ and $\varrho'^{\mu\nu}$, respectively (see \cite{landau_quantum_electrodynamics_1982}, \S~65):
\begin{equation}
	\varrho^{\mu\nu} = \sum_{i,j=1,2} \rho_{ij} e_i^\mu e_j^\nu, \quad \varrho'^{\mu\nu} = \sum_{i,j=1,2} \rho'_{ij} e_i^\mu e_j^\nu,
\end{equation}
where
\begin{gather}
	\label{eq:basis_landau}
	e_1^\mu = \frac{N^\mu}{\sqrt{-N^2}},\quad e_2^\mu = \frac{P^\mu}{\sqrt{-P^2}},
\end{gather} 
\begin{equation}
\begin{aligned}
	P^\mu &= (g^{\mu\nu} - \nfrac{K^\mu K^\nu}{K^2})  (p+p')_\nu,\\
	N^\mu &= \eps^{\mu\nu\rho\sigma} P_\nu Q_\rho K_\sigma,
\end{aligned}
\end{equation}
with $K^\mu = k^\mu + q^\mu$ and $Q^\mu = q^\mu - k^\mu$. We introduce the Stokes vectors $\boldsymbol{\xi} = (\xi_1, \xi_2, \xi_3)$ and $\boldsymbol{\xi}' = (\xi'_1, \xi'_2, \xi'_3)$ via
\begin{equation}
	\label{eq:density_matices}
	\rho = \frac{1}{2}(\II + \boldsymbol{\xi}\boldsymbol{\sigma}), \quad
	\rho' = \frac{1}{2}(\II + \boldsymbol{\xi'}\boldsymbol{\sigma}),
\end{equation}
where the following representation for the Pauli matrices is used \cite{landau_quantum_electrodynamics_1982}:
\begin{equation}
\sigma_1 = \begin{pmatrix} 0 & 1\\ 1 & 0\end{pmatrix},
\quad
\sigma_2  = \begin{pmatrix} 0 & -i\\ i & \phantom{-}0\end{pmatrix},
\quad
\sigma_3  = \begin{pmatrix} 1 & \phantom{-}0\\ 0 & -1\end{pmatrix}.
\end{equation}

Note that $\xi_1$ and $\xi_3$ correspond to linear polarization as $\varrho^{\mu \nu} = e^\mu e^{*\nu}$ with $e^\mu = \cos(\varphi_\gamma) e_1^\mu + \sin(\varphi_\gamma) e_2^\mu$ implies $\xi_1 = \sin(2\varphi_\gamma)$, $\xi_2 = 0$,  and $\xi_3 = \cos(2\varphi_\gamma)$ \{$\varphi_\gamma$ is an arbitrary azimuth angle $\in [0, 2\pi)$\}; whereas $\xi_2$ corresponds to circular polarization as $e^\mu = \nfrac{(e_1^\mu \pm i e_2^\mu)}{\sqrt{2}}$ implies $\xi_1=0$, $\xi_2 = \pm 1$, and $\xi_3=0$.

Using the above notation, the differential cross section for Compton scattering reads \cite{landau_quantum_electrodynamics_1982}
\begin{equation}
	\label{eq:compton_cross_section}
	\dd \sigma\indices*{_{\textrm{C}}} = \frac{1}{16\pi^2} |M_{fi}|^2 \frac{\omega^2 \dd \Omega}{m^4x^2},
\end{equation}
where
\begin{equation}
	\label{eq:compton_mfi_squared}
	\begin{aligned}
		|M_{fi}|^2 = 16 \pi^2 r_e^2 m^2 \big[ F_0 &+ F_3 \left( \xi_3 + \xi'_3 \right) + F_{11} \xi_1 \xi'_1\\
		&+ F_{22} \xi_2 \xi'_2 + F_{33} \xi_3 \xi'_3 \big],
	\end{aligned}
\end{equation}
\begin{equation}
	\begin{gathered}
		F_0 = V - F_3, \quad
		F_3 = - (U^2 + 2U),\\
		F_{11} = 2(1 + U), \quad
		F_{22} = V (1 + U),\\
		F_{33} = 2 - F_3,
	\end{gathered}
\end{equation}
$\dd \Omega = \sin\theta \dd \theta \dd \varphi$ is the solid angle for the scattered photon, i.e., $q^\mu = \omega(1, \cos\varphi\sin\theta, \sin\varphi\sin\theta, \cos\theta)$, $U = 2/x - 2/y$, $V = x/y + y/x$, and
\begin{equation}
	x = \frac{2pk}{m^2} = \frac{2\epsilon\omega_\text{bs}}{m^2}(1 + \beta),\quad
	y = \frac{2pq}{m^2} = \frac{2\epsilon\omega}{m^2}(1 + \beta\cos\theta).
\end{equation}
\begin{table*}[bt]
\begin{center}
\begin{tabular}{cccccccccc}
\toprule
& $\omega_\textrm{L}$~[eV] & $\mathcal{E}$~[J] & $\Delta t$~[fs] & $I$~[W/cm$^2$] & PRR~[Hz] & $\xi$ & $\chi$ & $\xi N$ & $\Delta\phi$\\
\midrule
Apollon F1~\cite{papadopoulos_apollon_2016} & $1.55^*$ & 150 & 15 & $10^{23*}$\hspace*{-3.8pt} & $1/60 $ & 150 & $\phantom{0}1.8 \times \omega [\unit{GeV}]$ & 860 & 30\\
ELI-NP (x2)~\cite{negoita_eli-np_2016,turcu_high_2016} & 1.55\hspace*{4.5pt} & 250 & 25 & $10^{23}$ & 1/60 & 150 & $\phantom{0}1.8 \times \omega [\unit{GeV}]$ & 1400 & 50\\
ELI-Beamlines L4~\cite{rus_eli-beamlines_laser_systems_2013,le_garrec_eli-beamlines_2014,eli-beamlines_lasers_website} & $1.55^*$ & 1500 & 150 & $10^{22}$ & 1/60 & 50 & $0.57 \times \omega [\unit{GeV}]$ & 2700 & 300\\
\bottomrule
\end{tabular}
\caption{\label{tab:facilities}\footnotesize Laser parameters which are considered in the numerical calculations (parameters with a star are not explicitly confirmed in the cited references). Note that ELI-NP hosts two lasers with the designated parameters.}
\end{center}
\end{table*}
\hspace*{-0.12cm}The energy $\omega$ of the final photon is determined via four-momentum conservation $p^\mu + k^\mu = p'^\mu + q^\mu$ and is given by
\begin{equation}
	\label{eq:compton_photon_energy}
	\omega = \frac{(1 + \beta)\epsilon\omega_\text{bs}}{\epsilon + \omega_\text{bs} - (\epsilon\beta - \omega_\text{bs})\cos\theta},
\end{equation}
where $\beta = |\boldsymbol{p}|/\epsilon$. Correspondingly, the highest energy is obtained for perfect backscattering ($\theta=0$):
\begin{gather}
	\omega_{\text{max}} = \frac{(1+\beta)^2\eps^2\omega_\text{bs}}{m^2 + 2(1+\beta)\eps\omega_\text{bs}} \approx \frac{4\eps^2\omega_\text{bs}}{m^2 + 4\eps\omega_\text{bs}}
\end{gather}
(the last relation holds for ultrelativistic electrons). We assume that in the experiment the monochromator selects photons scattered by angles $\varphi \in (0, 2\pi)$ and $\theta \in (0, \theta_{\textrm{max}})$, where $\theta_{\textrm{max}} \ll 1$. The total cross section (averaged over the initial and summed over the final photon polarization) for those photons is 
\begin{equation}
	\label{eq:compton_cross_section_integrated}
	\sigma_{\textrm{bs}} = \frac{4\pi r_e^2}{m^2x^2} \int\displaylimits_0^{\theta_{\textrm{max}}} \dd \theta\; \omega^2F_0 \sin\theta.
\end{equation}

In order to consider polarization effects we first note that [see \eqref{eq:basis_landau}]
\begin{equation}
	\label{eq:basis_landau_head-on}
	\begin{aligned}
		e_1^\mu &= (0,\sin\varphi,-\cos\varphi,0),\\
		e_2^\mu &= -[\tan(\nfrac{\theta}{2}),\cos\varphi,\sin\varphi,-\tan(\nfrac{\theta}{2})].
	\end{aligned}
\end{equation}
Therefore, the Stokes parameters $\xi_i$ and $\xi'_i$ [see \eqref{eq:density_matices}] implicitly depend on $\varphi$. We eliminate this dependence (to leading order in $\theta \ll 1$) by introducing another basis $\tilde{e}_i^\mu$ ($i=1,2$) which is given by
\begin{gather}
\tilde{e}_i^\mu = \sum_{j=1,2} R_{ij}(\varphi) e_j^\mu,
\quad
R(\varphi)
=
\begin{pmatrix}
\phantom{+} \cos\varphi & \sin\varphi\\
         -  \sin\varphi & \cos\varphi
\end{pmatrix},
\end{gather}
such that 
\begin{equation}
	\label{eq:basis_fixed}
	\begin{aligned}
		\tilde{e}_1^\mu(\theta=0) &= -\Lambda\indices*{_2^\mu} = (0,0,-1,0),\\
		\tilde{e}_2^\mu(\theta=0) &= -\Lambda\indices*{_1^\mu} = (0,-1,0,0).
	\end{aligned}
\end{equation}
We denote the Stokes parameters for the initial beam and the state selected by the detector in the new basis by $S\indices*{_{\textrm{bs},i}}$ and $S'_{\textrm{bs},i}$, respectively. They are related to $\xi_i$ and $\xi_i'$ via
\begin{gather}
	\begin{pmatrix}
		\xi\indices*{_{1}}\\
		\xi\indices*{_{3}}
	\end{pmatrix}
	= R(2\varphi) 
	\begin{pmatrix}
		S\indices*{_{\textrm{bs},1}}\\
		S\indices*{_{\textrm{bs},3}}
	\end{pmatrix},
	\begin{pmatrix}
		\xi_{1}'\\
		\xi_{3}'
	\end{pmatrix}
	= R(2\varphi) 
	\begin{pmatrix}
		S'_{\textrm{bs},1}\\
		S'_{\textrm{bs},3}
	\end{pmatrix},
\end{gather}
and $\xi_2 = S\indices*{_{\textrm{bs},2}}$, $\xi_2' = S'_{\textrm{bs},2}$.

In order to determine the Stokes parameters $S\indices*{_i^{(0)}}$ of the photon beam, which enters the strong laser pulse, we set $\theta = 0$ in the basis $\tilde{e}_i^\mu$ [$i=1,2$; see \eqref{eq:basis_fixed}] as $\theta \ll 1$ for all selected photons, and integrate the cross section [see \eqref{eq:compton_cross_section}] over $\varphi$. Finally, we obtain that (see \cite{landau_quantum_electrodynamics_1982}, \S~65, 87)
\begin{gather}
	\label{eq:stokes_averaged}
	\begin{gathered}
		S\indices*{_{1}^{(0)}}  = \frac{F_{11} + F_{33}}{2F_0} S\indices*{_{\textrm{bs},1}},
		\quad
		S\indices*{_{2}^{(0)}}  = \frac{F_{22}}{F_0} S\indices*{_{\textrm{bs},2}},
		\\
		S\indices*{_{3}^{(0)}}  = \frac{F_{11} + F_{33}}{2F_0} S\indices*{_{\textrm{bs},3}}.
	\end{gathered}
\end{gather}
Note that for $\theta=0$ we obtain $\nfrac{(F_{11} + F_{33})}{2F_0} = 0$ and $\nfrac{F_{22}}{F_0} = -1$. In the calculations we assume that the laser beam, employed for backscattering, is right-handed circularly polarized, i.e., $S\indices*{_{\textrm{bs},1}} = S\indices*{_{\textrm{bs},3}} = 0$, $S\indices*{_{\textrm{bs},2}} = 1$. Therefore, $S\indices*{_{1}^{(0)}} = S\indices*{_{3}^{(0)}} = 0$, $S\indices*{_{2}^{(0)}} \approx -1$ for small $\theta_\textrm{max}$. In order to obtain a highly polarized beam we choose $\theta_\textrm{max}$ such that $|F_{22}/F_0| > 0.999$ for all selected photons. The parameters and cross sections for the three facilities, which are considered in the main text, are shown in Table~\ref{tab:compton}.

\section{Pair production in a Coulomb field}
\label{sec:pair}

The cross section of electron-positron photoproduction by a photon with energy $\omega \gg m$ colliding with an atom (charge number $Z$) is given by [see Eq.~(10.3) of \cite{olsen_pair_production_1959}]
\begin{gather}
	\label{eq:ppcrosssection}
	\dd \sigma_{\textrm{pp}} = \frac{d\varphi}{2\pi} \, [\sigma_0 + \sigma_1 (2|\hat{\spvec{u}}\spvec{e}_\gamma|^2 -1)],
\end{gather}
where
\begin{multline}
	\label{eq:ppc_sigma_zero}
	\sigma_0 = 2\frac{Z^2 \alpha r_e^2}{\omega^3} \int\displaylimits_m^{\omega-m} d\eps \int\displaylimits_{m^2/\epsilon^2}^{1} d\zeta  
\\
\Big\{ (\eps^2+\eps'^2) (3 + 2\Gamma) 
+ 
2\eps\eps' \big[1+4\boldsymbol{u}^2\zeta^2\Gamma \big] \Big\}
\end{multline}
and
\begin{gather}
	\label{eq:ppc_sigma_one}
	\sigma_1 = 2\frac{Z^2 \alpha r_e^2}{\omega^3} \int\displaylimits_m^{\omega-m} d\eps \int\displaylimits_{m^2/\epsilon^2}^{1} d\zeta \, 8\eps\eps'\boldsymbol{u}^2\zeta^2\Gamma
\end{gather}
(we have summed over the spin states of the produced electron-positron pair). Here, $\spvec{p}$ denotes the electron momentum, $\eps = \sqrt{m^2 + \boldsymbol{p}^2}$ and $\eps' = \omega - \eps$ are the energy of the produced electron and positron, respectively; $\spvec{q} = |\spvec{q}|\spvec{e}_z$ ($|\spvec{q}| = \omega$) and $\spvec{e}_\gamma$ denote the momentum and the polarization vector of the incoming photon, respectively, and $\spvec{u}$ is the component of $\spvec{p}$ (scaled by $m$) perpendicular to $\spvec{q}$, it is defined as $\spvec{u} = [\spvec{p} - \hat{\spvec{q}}(\hat{\spvec{q}}\spvec{p})]/m$ ($\hat{\spvec{q}} = \spvec{q}/\omega$). In the frame we consider, $\spvec{u} = \{\spvec{u}_x,\spvec{u}_y\} = |\spvec{u}| \{\cos\varphi,\sin\varphi\}$ ($\hat{\spvec{u}} = \nfrac{\spvec{u}}{|\spvec{u}|}$). Furthermore, $\zeta = \nfrac{1}{(1 + \spvec{u}^2)}$ and
\begin{gather}
\Gamma = \ln (\nfrac{1}{\delta}) - 2 - f(Z) + \mathcal{F}(\nfrac{\delta}{\zeta}),
\end{gather}
where $\delta = \nfrac{m\omega}{(2\eps\eps')}$,
\begin{gather}
f(Z) = (Z\alpha)^2 \sum_{n=1}^{\infty} \frac{1}{n[n^2 + (Z\alpha)^2]}.
\end{gather}
\begin{table}
\begin{center}
\begin{tabular}{ccccc}
\toprule
& $\epsilon$~[GeV] & $\theta_\textrm{max}$~[rad] & $\sigma_{\text{bs}} [r_e^2]$ & $\omega$~[MeV] \\
\midrule
Apollon/ELI-NP & 2.5 & $3.0 \times 10^{-5}$ & $0.232$ & $140$ \\
ELI-Beamlines & 4.5 & $1.6 \times 10^{-5}$ & $0.197$ & $430$ \\
\bottomrule
\end{tabular}
\caption{\label{tab:compton}\footnotesize Parameters for Compton backscattering at the three considered facilities for measuring vacuum birefringence. Here, $\epsilon$ denotes the electron energy, $\theta_\textrm{max}$ is the selected maximal scattering angle [chosen such that $|F_{22}/F_0| > 0.999$], $\sigma_{\text{bs}}$ is the cross section [see \eqref{eq:compton_cross_section_integrated}], and $\omega$ is the final photon energy [see \eqref{eq:compton_photon_energy}]. We choose $\epsilon$ such that $\chi = 0.25$. The final photon energy $\omega$ differs by less than $2\%$ from the given value in the range $0 \leqslant \theta \leqslant \theta_\textrm{max}$.}
\end{center}
\end{table}
\hspace*{-0.12cm}Here, we employ the Thomas-Fermi model with Moli\`ere parametrization, i.e., the screening term is given by
\begin{multline}
\mathcal{F}(\nfrac{\delta}{\zeta}) = - \frac{1}{2} \sum_{i=1}^3 \alpha_i^2 \ln(1+B_i) 
\\ 
+ \sum_{\scriptsize\begin{gathered}i,j = 1\\[-5pt] i\neq j \end{gathered}}^3 \alpha_i \alpha_j \Big[ \frac{1+B_j}{B_i-B_j} \ln (1 + B_j) + \frac{1}{2} \Big]
\end{multline}
with $B_i = (\nfrac{\beta_i \zeta}{\delta})^2$, $\beta_i = (\nfrac{Z^{\nfrac{1}{3}}}{121}) b_i$ and
\begin{gather}
\begin{aligned}
	\alpha_1 &= 0.1,& \alpha_2 &= 0.55,& \alpha_3 &= 0.35,&
\\
	b_1 &= 6.0,& b_2 &= 1.2,& b_3 &= 0.3.&
\end{aligned}
\end{gather}

We rewrite the cross section given in \eqref{eq:ppcrosssection} as
\begin{equation}
	\label{eq:ppc_cross_section_e_outside}
	\dd \sigma_{\textrm{pp}} = \frac{d\varphi}{2\pi} \, \sum_{i,j=1}^{3} [\sigma_0 \delta^{ij} + \sigma_1 (2\hat{u}^i\hat{u}^j - \delta^{ij})] e^i_\gamma e^{*j}_\gamma.
\end{equation}
Furthermore, we introduce the density matrix $\rho$ and the Stokes vector $S=\{S_0,\spvec{S}\}$ for the incoming photons as
\begin{gather}
	\label{eq:ppc_polarization_replacement}
	e^i_\gamma e^{*j}_\gamma \to \sum_{a,b=1,2} e^i_{a} e^j_{b}  \rho_{ab},
	\quad
	\rho = \frac{1}{2} (S_0 \II + \spvec{S} \spvec{\sigma}),
\end{gather}
where $\spvec{e}_{1} = \spvec{e}_x = (1,0,0)$, $\spvec{e}_{2} = \spvec{e}_y = (0,1,0)$ [$\Lambda\indices{_1^\mu} = (0, \spvec{e}_x)$, $\Lambda\indices{_2^\mu} = (0, \spvec{e}_y)$].
Combining Eqs.~(\ref{eq:ppc_cross_section_e_outside}) and~(\ref{eq:ppc_polarization_replacement}) we obtain for the pair production cross section \cite{kelner_methods_1975}:
\begin{gather}
	\label{eq:pp_cross_section_final}
	\dd \sigma_{\textrm{pp}} = \frac{d\varphi}{2\pi} \, \big\{\sigma_0 S_0 + \sigma_1 [S_1 \sin(2\varphi) + S_3 \cos(2\varphi)] \big\}.
\end{gather}

Note that the cross section given by Eqs. (\ref{eq:ppc_sigma_zero}), (\ref{eq:ppc_sigma_one}), and (\ref{eq:pp_cross_section_final}) neglects electron-induced pair production and inelastic contributions. In the numerical calculations we assume tungsten ($Z=74$) as conversion material, therefore both effects are subdominant ($Z$ vs. $Z^2$ scaling)~\cite{tsai_pair_production_1974}. Moreover, most of the pairs are produced near the forward direction such that we can neglect the nuclear form factors~\cite{tsai_pair_production_1974}.

The cross sections of pair production for the parameters considered in the main text are shown in Table~\ref{tab:ppc}.

	\begin{table}[bt]
\begin{center}
\begin{tabular}{ccccc}
\toprule
& $\omega$~[MeV] & $\sigma_0 [r_e^2]$ & $\sigma_1 [r_e^2]$ & $\sigma_1 / \sigma_0$ \\
\midrule
Apollon/ELI-NP & 140 & 344 & 26.7 & 0.078 \\
ELI-Beamlines & 430 & 393 & 31.0 & 0.079 \\
\bottomrule
\end{tabular}
\caption{\label{tab:ppc}\footnotesize Pair production cross sections in tungsten ($Z = 74$) for the probe photon energies shown in Table~\ref{tab:compton}. The cross section $\sigma_0$ represents the unpolarized part [see \eqref{eq:ppc_sigma_zero}], whereas $\sigma_1$ determines the significance of polarization effects [see \eqref{eq:ppc_sigma_one}]. Note that the observables (see the main text of the paper) depend only on the ratio $\sigma_1 / \sigma_0$.}
\end{center}
\end{table}

\vspace{-0.1cm}
\section{Statistical analysis}
\label{sec:statistics}
\vspace{-0.1cm}

The observables introduced in the main text are asymmetries of the type
\begin{gather}
\label{eq:statistics_asym}
R = \frac{N_A - N_B}{N_A + N_B},
\end{gather}
where $N_A$ and $N_B$ are experimentally measured numbers of events.

We describe the experiment in the following way: with probabilities $p_A$ and $p_B$ a probe photon decays inside the detector such that the produced pair contributes to $N_A$ and $N_B$, respectively, and the probability $p_C = 1-p_A-p_B$ accounts for all other possibilities (the photon decays inside the strong laser pulse, passes through the detector, or the produced pair is detected out of the range corresponding to $N_A$ and $N_B$). Therefore, the two random variables $N_A$ and $N_B$ are distributed according to a multinomial distribution \cite{riley_mathematical_2006,james_statistical_2006}. Their expectation values are given by $\expectation{N_A} = p_A N_\gamma$ and $\expectation{N_B} = p_B N_\gamma$, respectively, where $N_\gamma$ denotes the number of gamma photons generated via Compton backscattering. The standard deviations are given by $\Delta N_A = \sqrt{N_\gamma p_A (1 - p_A)}$ and $\Delta N_B = \sqrt{N_\gamma p_B (1 - p_B)}$, respectively.

Assuming that the number of events counted is large we approximate the expectation value of the asymmetry defined in \eqref{eq:statistics_asym} by \cite{riley_mathematical_2006,james_statistical_2006,ku_notes_1966}
\begin{equation}
	\label{eq:asymmetry_expectation}
	\expectation{R} = \frac{\expectation{N_A} - \expectation{N_B}}{\expectation{N_A} + \expectation{N_B}}
\end{equation}
and the variance by \cite{riley_mathematical_2006,james_statistical_2006,ku_notes_1966}
\begin{multline}
	\label{eq:asymmetry_variance}
	(\Delta R)^2 = \left( \frac{\partial R}{\partial \expectation{N_A}} \Delta N_A \right)^2 + \left( \frac{\partial R}{\partial \expectation{N_B}} \Delta N_B \right)^2 \\+ 2 \left( \frac{\partial R}{\partial \expectation{N_A}} \right) \left( \frac{\partial R}{\partial \expectation{N_B}} \right) \cov[N_A,N_B],
\end{multline}
where
\begin{equation}
	\frac{\partial R}{\partial \expectation{N_i}} = \left. \frac{\partial R}{\partial N_i} \right|_{N_j=\expectation{N_j}} \quad\quad (i,j = A,B)
\end{equation}
and $\cov[N_A,N_B] = - p_A p_B N_\gamma$. Using Eqs. (\ref{eq:asymmetry_expectation}) and (\ref{eq:asymmetry_variance}) we find that
\begin{gather}
\expectation{R} = \frac{p_A - p_B}{p_A + p_B},
\quad
(\Delta R)^2 = \frac{1-\expectation{R}^2}{N_\gamma (p_A + p_B)}.
\end{gather}
Assuming $\expectation{R}^2 \ll 1$ we conclude that the standard deviation of the asymmetry is given by $\Delta R \approx \nfrac{1}{\sqrt{N_\gamma (p_A + p_B)}}$. The number of required incoming gamma photons is now obtained from the condition $\expectation{R} - \expectation{R_0}= n \Delta R$, where $\expectation{R_0} = 0$ is the expectation value of the asymmetry if vacuum birefringence/dichroism is absent.
We conclude that
\begin{equation}
	\vspace{-0.1cm}
	N_\gamma = \frac{n^2}{\expectation{R}^2(p_A + p_B)}.
\end{equation}

In order to obtain the expressions presented in the main text we take into account that for the considered setup the probabilities $p_A$ and $p_B$ are given by $p_{A/B} = n_z l \sigma_{A/B}$, where
\begin{equation}
	\vspace{-0.1cm}
	\sigma_{A/B} = \frac{2\beta}{\pi}S_0\sigma_0 \pm \frac{\sin(2\beta)}{\pi}S_i\sigma_1
\end{equation}
with $S_i = S_1$ and $S_i = S_3$ for the measurement of vacuum birefringence and dichroism, respectively.

\vspace{-0.2cm}
\section{LINAC-based experiment}
\label{sec:accelerator}
\vspace{-0.2cm}

In the main text we also consider a potential vacuum birefringence experiment which could be performed at a conventional linear accelerator (LINAC) if combined with a 1~PW laser (intensity $I = \unitfrac[10^{21}]{W}{cm^2}$). In order to achieve $\chi = 0.25$ probe photons with an energy of $\omega = \unit[1.4]{GeV}$ are required. They are obtainable via Compton backscattering off an $\unit[8.4]{GeV}$ electron beam ($\theta_\textrm{max} = \unit[7.6 \times 10^{-6}]{rad}$, $\sigma_\textrm{bs} = 0.135 r_e^2$, $\sigma_1/\sigma_0 = 0.077$). Such electron energies are/will be available, e.g., at the European XFEL (up to $\unit[17.5]{GeV}$)~\cite{euroxfel}, FACET-II (up to $\unit[10]{GeV}$)~\cite{facet-ii_website}, and SACLA (up to $\unit[8.5]{GeV}$)~\cite{yabashi_overview_2015}. All these accelerators operate with a repetition rate of at least $\unit[10]{Hz}$, therefore, we assume that they are combined with a $\unit[10]{Hz}$ laser having the same parameters as the L3 laser being installed at ELI-Beamlines ($\omega_\text{L} = \unit[1.55]{eV}$, $\mathcal{E} = \unit[30]{J}$, $\Delta t = \unit[30]{fs}$) \cite{eli-beamlines_lasers_website}. Furthermore, we assume $N_e = 10^8$. Note that electron bunches with $N_e = 10^9$ electrons and a beam spot area $\sim \unit[30]{\upmu{}m^2}$ are envisaged for FACET-II \cite{facet-ii_website}. If, instead, only gamma photons from $N_e = 10^7$ electrons hit the high-intensity region of the optical laser, the required measurement time is increased from 2 to 20 days.

Note that for FACET-II a hundred-TW-class laser ($\omega_\text{L} = \unit[1.55]{eV}$, PRR = 10~Hz) and $\eta = 10^{-3}$ could be sufficient. Assuming 10-GeV electrons, $N_e = 10^9$, $\theta_\textrm{max} = \unit[6 \times 10^{-6}]{rad}$ (i.e., $\omega = \unit[1.9]{GeV}$, $\sigma_\textrm{bs} = 0.113 r_e^2$, $\sigma_1/\sigma_0 = 0.077$) the measurement time is 3 hours if using a 200 TW laser (20~J in 100~fs, $I = \unitfrac[5 \times 10^{20}]{W}{cm^2}$) and 12 days if using a 100 TW laser (4~J in 35~fs, $I = \unitfrac[2.3 \times 10^{20}]{W}{cm^2}$).

\vspace{-0.3cm}